\documentclass[aps,preprint,epsfig,rotate]{revtex4}
\usepackage{graphicx}
\usepackage{bm}
\usepackage{epsfig}

\input epsf
\begin{document}
\title{On the hyperfine structures of the ground state(s) in the ${}^{6}$Li and ${}^{7}$Li atoms}

\author{Alexei M. Frolov}
\email[E--mail address: ]{afrolov@uwo.ca}

\affiliation{Department of Applied Mathematics \\
 University of Western Ontario, London, Ontario N6H 5B7, Canada}

\date{\today}

\begin{abstract}

Hyperfine structure of the ground $2^{2}S-$states of the three-electron atoms and ions is investigated. By using our recent numerical values for the 
doublet electron density at the atomic nucleus we determine the hyperfine structure of the ground (doublet) $2^{2}S-$state(s) in the ${}^{6}$Li and 
${}^{7}$Li atoms. Our predicted values (228.2058 $MHz$ and 803.5581 $MHz$, respectivly) agree well with the experimental values 228.20528(8) $MHz$ 
(${}^{6}$Li) and 803.50404(48) $MHz$ (${}^{7}$Li (R.G. Schlecht and D.W. McColm, Phys. Rev. \textbf{142}, 11 (1966))). The hyperfine structures of a 
number of lithium isotopes with short life-times, including ${}^{8}$Li, ${}^{9}$Li and ${}^{11}$Li atoms are also predicted. The same method is used 
to obtain the hyperfine structures of the three-electron ${}^{7}$Be$^{+}$ and ${}^{9}$Be$^{+}$ ions in their ground $2^{2}S-$states. Finally, we 
conclude that our approach can be generalized to describe the hyperfine structure in the triplet $n{}^{3}S-$states of the four-electron atoms and ions. 

\noindent 
PACS number(s): 32.10.Fn, 31.15.A- and 31.15.Ve

\noindent 
This manuscript will be published in the {\bf Letters to JETP}, {\bf 103}, issue 12, xxx (2016).

\end{abstract}

\maketitle
\newpage

In this short communication we report the results of our analysis of the hyperfine structure of the bound ${}^{2}S-$states in three-electron atomic systems. Our main 
goal in this study is to develop the new method which can be used for direct computations of the hyperfine structure and hyperfine structure splitting of the ground 
$2^{2}S-$state(s) in the ${}^{6}$Li and ${}^{7}$Li atoms. In the future we want to apply the same approach to determie the hyperfine structure of the bound triplet 
${}^{3}S-$states in four-electron atoms and ions. For three-electron atoms and ions the problem is formulated as follows. The ground $2^{2}S-$state(s) in such atomic 
system is a doublet (electron) state. The total electron spin in the doublet atomic $S-$states is not zero and, therefore, one can observe the effect of direct spin-spin 
interaction between spin of the outermost electron and non-zero spin of the central atomic nucleus. The arising hyperfine structure is of great interest in a number of 
areas of atomic and molecular physics as well as in astrophysics \cite{Tenn}.  

In general, the hyperfine structure of the bound $S-$state in an arbitrary few-electron atom/ion is described by the Fermi-Segr\'{e} formula (see, e.g., \cite{LLQ}) 
\begin{equation}
    E_{hf} = \frac{8 \pi \alpha^2}{3} \mu_B \mu_N g_e g_N \cdot \rho_D(0) \cdot \frac12 [F(F + 1) - I_N(I_N + 1) - S_e(S_e + 1)] \label{eq1}
\end{equation} 
where $E_{hf}$ are the corresponding energy levels, $\rho_D(0)$ is the electron density at the central atomic nucleus (see below), $S_{e}(S_{e} + 1), I_{N}(I_{N} + 1)$ and 
$F(F + 1)$ are the eigenvalues of the three following operators: ${\bf S}^{2}_{e}, {\bf I}^{2}_{N}$ and ${\bf F}^{2}$, respectively. Here and below the notation 
${\bf S}_{e}$ is the electron spin of the atom, ${\bf I}_{N}$ is the spin of the nucleus (here we assume that $\mid {\bf I}_{n} \mid > 0$) and ${\bf F}$ is the total angular 
momentum operator of the atom. For bound $S-$states the operator ${\bf F}$ is considered as the total spin of the atom, i.e. the sum of the electron and nuclear spins. Also, 
in Eq.(\ref{eq1}) the factor $\alpha \approx$ 7.297 352 568 $\cdot 10^{-3} \approx \frac{1}{137}$ is the dimensionless fine structure constant, $\mu_B = \frac12$ is the Bohr 
magneton (in atomic units) and $\mu_N = \mu_B \frac{m_e}{m_p}$, where $m_p$ = 1836.152 672 61 $m_e$, is the inverse ratio of the proton and electron masses. The electron 
gyromagnetic ratio $g_e$ = -2.002 231 930 437 18 \cite{CRC}, while $g_N = \frac{f_N}{I_N}$ where the values $f_N$ and $I_N$ are defined below. Finally, the formula for the 
energy levels of hyperfine structure $E_{hf}$ takes the form
\begin{eqnarray}
   E_{hf}(MHz) = -400.1012444432845 \cdot \Bigl(\frac{f_N}{I_N}\Bigr) \rho_D(0) \cdot [F(F + 1) - I_N(I_N + 1) - S_e(S_e + 1)] \label{eq2}
\end{eqnarray}
where the factor 6.579 683 920 61$\cdot 10^9$ ($MHz/a.u.$) is used to re-calculate the result from the atomic units to $MegaHertz$ which are more convenient for comparison with
experiments. In Eq.(\ref{eq2}) the factor $f_N$ is the actual magnetic momentum of the nucleus and $I_N$ is the nuclear spin. Both these values can be found in tables of nuclear 
properties (see, e.g., the corresponding Section in \cite{CRC}). 

Let us briefly discuss the electron density at the atomic nucleus, or in other words, the factor $\rho_D(0)$ in Eqs.(\ref{eq1}) and (\ref{eq2}). For one- and two-electron atoms 
and ions this value coincides with the total electron density at the atomic nucleus, i.e. $\rho(0) = \langle \delta({\bf r}_{eN}) \rangle$, where $\langle \delta({\bf r}_{eN}) 
\rangle$ is the expectation value of the atomic electron-nucleus delta-functions determined for the triplet $2^3S-$state \cite{BS}. Here and below, the index $e$ designates the 
electron, while another index $N$ means the atomicnucleus. However, if the number of bound electrons in atom/ion exceeds two, then such a definition of the electron density at 
the atomic nucleus does not works, since the two internal electrons form the closed $1s^2-$elecron shell (with zero spin) and the outermost electron(s) must penetrate this shell 
to produce the actual spin-spin interaction with the atomic nucleus. Therefore, in such cases the factor $\rho_D(0)$ must be chosen as the electron density of the doublet, 
triplet, etc, atomic electrons at the atomic nucleus. It is clear that the $\rho_D(0)$ value will be quite different from the total electron density at the nucleus defined above 
$\rho(0) = \langle \delta({\bf r}_{eN}) \rangle$. For three-electron atoms/ions this is the doublet electron density (at the atomic nucleus), or density of the outermost 
$2s-$electron at the atomic nucleus \cite{Fermi}. The general definition of the doublet electron density at the central atomic nucleus, i.e. the $\rho_D(0)$ value in 
Eq.(\ref{eq2}) can be given in the form $\rho_D(0) = \langle \sum^{N_e}_{i=1}\delta({\bf r}_{iN}) (\sigma_z)_i \rangle$, where $N_e$ is the total number of bound atomic 
electrons and $(\sigma_z)_i$ is the Pauli $\sigma_z$ matrix for $i$-th  electron. For electrons with spin-up ($\alpha$) and spin-down $(\beta$) functions one finds: $(\sigma_z) 
\alpha = \alpha$ and $(\sigma_z) \beta = -\beta$ (see, e.g., \cite{LLQ}). As follows from this definition of the electron doublet density $\rho_{D}(0)$ its value equals zero 
identically for any closed electron shell and for any singlet state of the atom. For the doublet states $\rho_D(0) \ne 0$ and its numerical value can be found only from accurate 
numerical computations of the ${}^{2}S-$states in three-electron atoms (see below). 
 
Below, we restrict ourselves to the cases of the ${}^{6}$Li and ${}^{7}$Li atoms for which we have $f_N = 0.8220473, I_N = 1^{+}$ and $f_N = -3.2564268, I_N = \Bigl(\frac32\Bigr)^{-}$, 
respectively \cite{CRC}. In these cases one finds for the total spin of the atom $F$ in the doublet $S-$states: $F = \frac12$, or $\frac32$ (for ${}^{6}$Li) and $F = 1$, or $2$ 
(for ${}^{7}$Li). With these values of $F$ one can easily determine the numerical values of the $G = F(F + 1) - I_N(I_N + 1) - S_e(S_e + 1)$ factor in Eq.(\ref{eq2}) for the 
${}^{6}$Li and ${}^{7}$Li atoms: $G$ = -2 and $G$ = 1 (for ${}^{6}$Li) and $G = -\frac52$ and $G = \frac32$ (for ${}^{7}$Li). Thus, we have six (2 + 4) energy levels of hyperfine 
structure in the ${}^{6}$Li atom and eight analogous levels (3 + 5) in the ${}^{7}$Li atom. The differnce between the energies of these levels is called the hyperfine structure 
splitting $\Delta E_{hf}$. In other words, $\Delta E_{hf} = \mid E_{hf}(F = \frac12) - E_{hf}(F = \frac32) \mid$ for the ${}^{6}$Li atom and $\Delta E_{hf} = \mid E_{hf}(F = 1) 
- E_{hf}(F = 2) \mid$ for the ${}^{7}$Li atom.
   
Numerical computations of the ground $2^2S-$state in the three-electron Li atom have been performed by applying the variational wave functions which are similar to the wave 
function used in \cite{Fro1}. Each of these variational wave functions contains two spin function $\chi^{(1)}_{S=\frac12} = \alpha \beta \alpha - \beta \alpha \alpha$ and 
$\chi^{(2)}_{S=\frac12} = 2 \alpha \alpha \beta - \beta \alpha \alpha - \alpha \beta \alpha$. The corresponding radial factors of the total wave functions have been approximated 
with the use of KT-variational expansion \cite{KT} in multi-dimensional gaussoids of the relative coordinates $r_{ij} = \mid {\bf r}_i - {\bf r}_j \mid$. For three-electron (or 
four-body) Li atom we used six-dimensional gaussoids which depend upon six relative coordinates $r_{12}, r_{13}, r_{23}, r_{14}, r_{24}, r_{34}$ \cite{Fro1}. In these notations 
the notations/indexes 1, 2, 3 designate three atomic electrons, while 4 means heavy atomic nucleus. The explicit form of our variational expansion is
\begin{eqnarray}
  \psi(L = 0; S = \frac12) = \sum^{N_A}_{i=1} C_i {\cal A}_{123} [\exp(-a_{ij} r^2_{ij}) \chi^{(1)}_{S=\frac12}] + 
  \sum^{N_B}_{i=1} G_i {\cal A}_{123} [\exp(-b_{ij} r^2_{ij}) \chi^{(2)}_{S=\frac12}] \; \; \; \label{eq3}
\end{eqnarray}
where ${\cal A}_{123}$ is the complete three-electron (or three-particle) anti-symmetrizer, $C_i$ (and $G_i$) are the linear variational coefficients of the variational function, while 
$\alpha_{ij}$,  where $(ij)$ = (12), (13), $\ldots$, (34), are the six non-linear parameters for three-electron atomic systems.  Analogously, the notation $b_{ij}$ stands for other six 
non-linear parameters which must also be varied (independently of $a_{ij}$) in calculations. By varying these non-linear parameters in the wave functions one can approximate the actual 
wave function(s) to high accuracy. In our calculations we used wave functions with 700, 1000 and 1400 basis functions. The wave functions with 700 basis radial functions produces 
-7.478007583 $a.u.$ for the total energy of the ground $2^2S-$state of the Li atom. Note that with this wave function we have also found that $\langle \delta({\bf r}_{eN}) \rangle = 
4.607884, \langle \delta({\bf r}_{ee}) \rangle = 0.181641$ and $\rho_D(0)\approx 0.2312796$ $a.u.$ (all values are given in atomic units ($a.u.$)). A large number of other bound state 
properties of the Li atom in its ground state, computed to very good accuracy with the wave funtion(s) Eq.(\ref{eq3}), can be found in \cite{Fro1}. In general, the bound state properties 
computed with the trial wave function, Eq.(\ref{eq3}), are very close (or even better, see \cite{Fro1}) than the corresponding expectation values obtained for the ground state in the Li 
atom in \cite{King}. It should be mentioned that all these calculations have been performed for the lithium atom with infinitely heavy nucleus (${}^{\infty}$Li atom). 

Now, by using our numerical value of the doublet electron density $\rho_D(0)$ obtained for the ground $2^2S-$state of the ${}^{\infty}$Li atom (see above), we can predict the hyperfine 
structure of the ${}^{6}$Li and ${}^{7}$Li atoms in this state. The corresponding hyperfine enegy levels for these atoms can be found in Table I (in $MHz$). In general, our results from 
Table I coincide very well with the experimental data from \cite{Liatom}, where the hyperfine structure of the ground states in the ${}^{6}$Li and ${}^{7}$Li atoms was investigated. 
Indeed, the hyperfine structure splitings observed in \cite{Liatom} are: 228.20528(8) $MHz$ (${}^{6}$Li) and 803.50404(48) $MHz$ (${}^{7}$Li). The overall coincidence of our results and 
experimental data for the hyperfine structure spliting in the ${}^{6}$Li atom is significantly better than in the case of ${}^{7}$Li atom, where the actual difference exceeds 50 $kHz$. 
Nevertheless, we could not expect such a good agreement with experimental data for our simple wave function which has been constructed for the model ${}^{\infty}$Li atom. The same method 
can be used to predict the hyperfine structure in an arbitrary three-electron Li-like ion which is located in the ground $2^{3}S-$state. For instance, for the ${}^{9}$Be$^{+}$ ion by using 
its nuclear data ($f_N = -1.177432, I_N = \Bigl(\frac32\Bigr)^{-}$) and our expectation value for the doublet electron density at the atomic nucleus ($\rho_D(0) \approx$ 0.993581) one 
finds that the hyperfine structure of the ground state of ${}^{9}$Be$^{+}$ ion is very similar to hyperfine structure of the ground state of the ${}^{7}$Li atom considered above. The 
hyperfine energy levels and hyperfine structure splitting between them can be found in Table I. The multiplicities of these levels of hyperfine structure can easily be predicted from 
similarity of the hyperfine structures of the ${}^{7}$Li atom and ${}^{9}$Be$^{+}$ ion. The hyperfine structure spliting, evaluated with our expectation value of the doublet density 
$\rho_D(0)$ computed for the ground $2^{3}S-$state of the Be$^{+}$ ion, is $\Delta_{hf} \approx$ 1248.1818 $MHz$. To the best of our knowledge this value has not been measured in 
experiments. 

For the triplet $2^{3}S$-state of the neutral ${}^{9}$Be atom situation is opposite, since it is difficult to perform direct and accurate computations of the hyperfine structure for this 
four-electron atomic system. In reality, it is hard to determine the accurate numerical value of the triplet electron density at the atomic nucleus $\rho_T(0)$ for four-electron atoms and 
ions. In contrast with the three-electron atomic systems discussed in this study any trial wave function of the Be-atom contains four times more spin states and in accurate computations 
we have to use the two spin functions: $\chi^{(1)}_{S=1} = \alpha \beta \alpha \alpha - \beta \alpha \alpha \alpha$ and $\chi^{(2)}_{S=1} = 2 \alpha \alpha \beta \alpha - \beta \alpha \alpha 
\alpha - \alpha \beta \alpha \alpha$. However, as follows from the results of this study the hyperfine structure of the ${}^{9}$Be atom can be investigated by using the new modification 
of our method to four-electron atoms/ions developed in this paper. This is the main goal for future studies. In conclusion, let us note that currently the hyperfine structure of the $P-$ 
and $D-$ states in lithium atoms is of increasing interest for many experimental and theoretical groups \cite{San} - \cite{Puch}. This can be explaind by a number of discrepancies between
recent theretical and experimental results (for more details, see, e.g., \cite{San} and references therein). In addition to this there is a disagreement between numerical results obtained 
by different theoretical groups. In this study we cannot discuss these interesting problems, since here we consider only the bound $S(L = 0)-$states in the lithium atom(s). Analysis of the
hyperfine structure splitting for the $S(L = 0)-$states and states with $L \ge 1$ in three-electron atoms/ions has a number of fundamental differences.

\newpage
\begin{table}[tbp]
   \caption{Predicted and observed hyperfine structures and hyperfine structure splittings in the ground ${}^{2}S-$state(s) in the ${}^{6}$Li and 
            ${}^{7}$Li atoms and ${}^{9}$Be$^{+}$ ion (in $MHz$). Experimental data have been taken from \cite{Liatom}.}
     \begin{center}
     \scalebox{0.95}{%
     \begin{tabular}{| c | c | c | c | c | c | c |}
      \hline\hline
  ${}^{6}$Li & $f_N$ & $I_N$ & $E_{hf}(1)$ & $E_{hf}(2)$ & $\Delta E_{hf}$ (predicted) & $\Delta E_{hf}$ (experiment) \\
          \hline    
  ${}^{6}$Li & 0.8220473 & $1^{+}$ & -76.0684 ($F = \frac32$) & 152.1372 ($F = \frac12$) & 228.2058 & 228.20528(8) \\
         \hline    
  ${}^{7}$Li & 3.2564268 & $\Bigl(\frac32\Bigr)^{-}$ & -502.2238 ($F$ = 1) & 301.3343 ($F$ = 2) & 803.5581 & 803.50404(48) \\ 
         \hline    
  ${}^{8}$Li & 1.653560 & $2^{+}$ & -306.0252 ($F = \frac52$) & 229.5189 ($F = \frac32$) & 535.5441 & -------- \\
         \hline    
  ${}^{9}$Li & 3.43678 & $\Bigl(\frac32\Bigr)^{-}$ & -530.0388 ($F$ = 1) & 318.0233 ($F$ = 2) & 848.0621 & -------- \\ 
         \hline    
  ${}^{11}$Li & 1.653560 & $\Bigl(\frac32\Bigr)^{-}$ & -566.1925 ($F$ = 1) & 339.7154 ($F$ = 2) & 905.9078 & -------- \\ 
          \hline\hline
  ${}^{7}$Be$^{+}$ & -1.39928  & $\Bigl(\frac32\Bigr)^{-}$ & -927.1002 ($F = 1$) & 556.2601 ($F$ = 2) & 1483.3603 & -------- \\

  ${}^{9}$Be$^{+}$ & -1.177432 & $\Bigl(\frac32\Bigr)^{-}$ & -780.1136 ($F = 1$) & 468.0681 ($F$ = 2) & 1248.1818 & -------- \\   
    \hline\hline
  \end{tabular}}
  \end{center}
  \end{table}
\end{document}